\documentclass[proceedings]{JHEP} % 10pt is ignored!
\usepackage{epsfig}         % please use epsfig.
\def\beqa{\begin{eqnarray}}
\def\eeqa{\end{eqnarray}}

\def\slashchar#1{\setbox0=\hbox{$#1$}           % set a box for #1
   \dimen0=\wd0                                 % and get its size
   \setbox1=\hbox{/} \dimen1=\wd1               % get size of /
   \ifdim\dimen0>\dimen1                        % #1 is bigger
      \rlap{\hbox to \dimen0{\hfil/\hfil}}      % so center / in box
      #1                                        % and print #1
   \else                                        % / is bigger
      \rlap{\hbox to \dimen1{\hfil$#1$\hfil}}   % so center #1
      /                                         % and print /
   \fi}                                         %
\title{Inclusive and exclusive decays of doubly heavy baryons}
\author{A.I.Onishchenko\\
        Institute for Theoretical and Experimental Physics,\\
        B.Cheremushkinskaja, 25, Moscow,117259 Russia\\
        Email: \email{onischen@heron.itep.ru}}
\conference{Heavy Quark Physics 5, Dubna, Russia, 6-8 April 2000}
\abstract{In this paper we give a short review of the recently
obtained results on inclusive and exclusive decays of doubly heavy
baryons.} 
\begin{document}

\section{Introduction}

Recently there was a big progress in our theoretical understanding
of the physics of doubly heavy baryons.  First, the production
cross sections of doubly heavy baryons in hadron collisions at
high energies of colliders and in fixed target experiments were
calculated with the use of perturbative QCD for the hard processes
and factorization hypothesis to account for the nonperturbative
binding of heavy quarks inside the doubly heavy baryons
\cite{prod}. Second, the lifetimes and branching fractions of some
inclusive decay modes were evaluated in the Operator Product
Expansion combined with the effective theory of heavy quarks
\cite{ltime,DHD}. Third, the families of doubly heavy baryons,
which contain a set of narrow excited levels in addition to the
basic state, were described in the framework of potential models
\cite{pot}. The picture of spectra, obtained in this analysis, is
very similar to that of heavy quarkonia. Fourth, the QCD and NRQCD
sum rules \cite{SVZ} were explored for the two-point baryonic
correlators in order to calculate the masses and couplings of
doubly heavy baryons \cite{QCDsr,DHSR1,DHSR2}. And fifth, there
are papers, where exclusive semileptonic and some nonleptonic
decay modes of doubly heavy baryons in the framework of
Bethe-Salpeter, NRQCD sum rules and potential models were analyzed
\cite{Lozano,Guo,Onish}. In the present talk we will concentrate
on the developments in the description of inclusive and exclusive
decays of the mentioned hadrons. In what follows, we will present
the results of OPE approach on lifetimes and the results of
three-point NRQCD sum rules on semileptonic and various
nonleptonic decay modes of doubly heavy baryons.

\section{Inclusive decays in OPE}

In the first part of this review we give a short description of
the OPE framework used to calculate lifetimes of doubly heavy
baryons and present numerical predictions for their values. Here
we also comment on relative contributions of spectator and
nonspectator effects to estimated lifetimes.

\subsection{OPE framework for lifetimes}

Let us describe the calculation framework for the lifetimes of
doubly heavy baryons on the concrete example of
$\Xi_{bc}^{\diamond }$ baryons. The optical theorem along with the
hypothesis of integral quark-hadron duality, leads us to a
relation between the total decay width of heavy quark and the
imaginary part of its forward scattering amplitude. This
relationship, applied to the $\Xi_{bc}^{\diamond
}$-baryon\footnote{Here $\diamond $ denotes electrical charge of
$\Xi_{bc}^{\diamond}$-baryon} total decay width
$\Gamma_{\Xi_{bc}^{\diamond }}$, can be written down as:
\begin{equation}
\Gamma_{\Xi_{bc}^{\diamond}} =
\frac{1}{2M_{\Xi_{bc}^{\diamond}}}\langle\Xi_{bc}^{\diamond
}|{\cal T}|\Xi_{bc}^{\diamond }\rangle,
\end{equation}
with the transition operator $\cal T$:
\begin{equation}
{\cal T} = {\cal I}m\int d^4x \{\hat TH_{eff}(x)H_{eff}(0)\},
\end{equation}
where the effective lagrangian of weak interactions $H_{eff}$, for
example, in the case of nonleptonic decays and at the
characteristic hadron energies is given by
\begin{eqnarray}
H_{eff} &=& \frac{G_F}{2\sqrt
2}V_{q_3q_4}V_{q_1q_2}^{*}[C_{+}(\mu)O_{+} + \nonumber \\ &&
C_{-}(\mu)O_{-}] + h.c. \nonumber
\end{eqnarray}
where
\begin{eqnarray}
O_{\pm} &=& [\bar
q_{1\alpha}\gamma_{\nu}(1-\gamma_5)q_{2\beta}][\bar
q_{3\gamma}\gamma^{\nu}(1-\gamma_5)q_{4\delta}]\times\nonumber \\
&& (\delta_{\alpha\beta}\delta_{
\gamma\delta}\pm\delta_{\alpha\delta}\delta_{\gamma\beta})\nonumber,
\end{eqnarray} and $$ C_+ = \left [\frac{\alpha_s(M_W)}{\alpha_s(\mu)}\right
]^{\frac{6}{33-2f}}, \quad C_- = \left
[\frac{\alpha_s(M_W)}{\alpha_s(\mu)}\right
]^{\frac{-12}{33-2f}},\\ $$ where f is the number of flavors and
$\{\alpha,\beta,\gamma,\delta \}$ run over the color indices.

As the energy release in heavy quarks decays is large, we may
benefit from the Operator Product Expansion (OPE) for the
transition operator
\begin{eqnarray}
{\cal T} &=& \sum_{i=1}^2\{C_1(\mu )\bar
Q^iQ^i+\frac{1}{m_{Q^i}^2}C_2(\mu )\bar Q^ig\sigma_{\mu\nu
}G^{\mu\nu }Q^i\nonumber \\ && + \frac{1}{m_{Q^i}^3}O(1)\}.
\end{eqnarray}
Performing the above expansion, we obtain a series of operators,
classified according to their dimensions. The contributions of
these operators to the total decay width of the baryon under
consideration have a simple physical interpretation:
\begin{itemize}
\item dimension 3: $\bar QQ$, this operator  represents the
contribution of spectator heavy quark decay.
\item dimension 4: removed by the equations of motion.
\item dimension 5: $Q_{GQ} = \bar Qg\sigma_{\mu\nu }G^{\mu\nu }Q$,
represents chromomagnetic interaction of the decaying quark with
other heavy quark as well as with the light quark.
\item dimension 6: $Q_{2Q2q} = \bar Q\Gamma q\bar q\Gamma^{'}Q$,
the operators of this kind correspond to nonspectator effects, the
most important of which are Pauli interference and weak scattering
\end{itemize}

\setlength{\unitlength}{1mm}
\begin{figure}[th]
\vspace*{-9mm}
\begin{center}
\begin{picture}(400.,30.)
\hspace*{1.cm} \epsfxsize=6cm \epsfbox{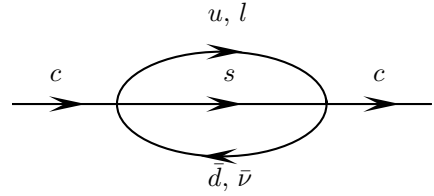} \put(-55,17){$c$}
\put(-12,17){$c$} \put(-32,17){$s$} \put(-34,25){$u$, $l$}
\put(-34,3){$\bar d$, $\bar \nu$}

\end{picture}
\end{center}
\vspace*{-8mm}
%{\hfill $z$\hspace*{8mm}}
\caption{The diagram of spectator contribution in the charmed
quark decays.} \label{fig1}
\end{figure}

Thus the transition operator can be written as
\begin{eqnarray}
{\cal T}_{\Xi_{bc}^{+}} &=& {\cal T}_{35b} + {\cal T}_{35c} +
{\cal T}_{6,PI}^{(1)} + {\cal T}_{6,WS}^{(1)}, \nonumber \\ {\cal
T}_{\Xi_{bc}^{0}} &=& {\cal T}_{35b} + {\cal T}_{35c} + {\cal
T}_{6,PI}^{(2)} + {\cal T}_{6,WS}^{(2)}. \nonumber
\end{eqnarray}
All contributions in the above expressions can be explicitly
calculated and, for example, the contribution of  dimension 3 and
5 operators in the case of $b$ - quark decay is given by the
following expression
\begin{eqnarray}
{\cal T}_{35b} &=& \Gamma_{b,spect}\bar bb -
\frac{\Gamma_{0b}}{m_b^2}[2P_{c1}+P_{c\tau 1}+\nonumber \\ &&
K_{ob}(P_{c1}+P_{cc1})+K_{2b}(P_{c2}+P_{cc2})]O_{Gb}, \nonumber
\end{eqnarray}
where
\begin{eqnarray}
\Gamma_{0b} = \frac{G_F^2m_b^5}{192\pi^3}|V_{cb}|^2,\nonumber
\end{eqnarray}
with $K_{0Q} = C_{-}^2+2C_{+}^2$, $K_{2Q} = 2(C_{+}^2-C_{-}^2)$,
\begin{eqnarray}
P_{c1} = (1-y)^4,\quad P_{c2} = (1-y)^3,\quad y =
\frac{m_c^2}{m_b^2}. \nonumber
\end{eqnarray}
Below, we have also written a characteristic nonspectator
contribution given by electroweak scattering of $b$ and $c$ -
quarks
\begin{eqnarray}
{\cal T}_{WS,bc} &=& \frac{G_F^2|V_{cb}|^2}{4\pi
}m_b^2(1+\frac{m_c}{m_b})^2(1-z_{+})^2\times \nonumber \\
&&[(C_{+}^2+C_{-}^2+\frac{1}{3}(1-k^{1/2})(C_{+}^2-C_{-}^2))\times\nonumber
\\ && (\bar b_i\gamma_{\alpha }(1-\gamma_5)b_i)(\bar c_j\gamma^{\alpha }(1-\gamma_5)c_j)
+\nonumber \\ && k^{1/2}(C_{+}^2-C_{-}^2)\times \nonumber \\ &&
(\bar b_i\gamma_{\alpha }(1-\gamma_5)b_j)(\bar
c_j\gamma^{\alpha}(1-\gamma_5)c_i) ],\nonumber
\end{eqnarray}
where
\begin{eqnarray}
z_{+} = \frac{m_c^2}{(m_b+m_c)^2},\quad k = \frac{\alpha_s (\mu
)}{\alpha_s (m_b+m_c)}. \nonumber
\end{eqnarray}
The hadronic matrix elements can be further estimated using
effective theories description of bound state dynamics of doubly
heavy baryons. Here we will not give the details of these
estimates and refer the interested reader for details to
\cite{ltime,DHD}. So, in the next subsection, we will go directly
to the numerical estimates of doubly heavy baryon lifetimes.

\setlength{\unitlength}{1mm}
\begin{figure}[th]
\vspace*{-9mm}
\begin{center}
\begin{picture}(400.,30.)
\hspace*{1.cm} \epsfxsize=6cm \epsfbox{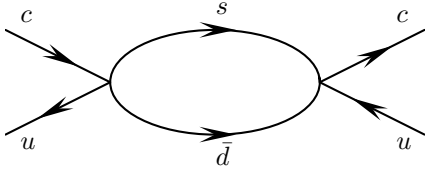} \put(-58,22){$c$}
\put(-58,5){$u$} \put(-8,22){$c$} \put(-8,5){$u$}
%\put(-50,24){$s$}
\put(-32,23){$s$} \put(-32,3){$\bar d$}

\end{picture}
\end{center}
\vspace*{-5mm} \caption{The diagram for the contribution of Pauli
interference in the decays of charmed quark for the $\Xi_{bc}^{+}$
baryon.} \label{fig2}
\end{figure}

\subsection{Numerical results}

Now we have already estimates for the lifetimes of all doubly
heavy baryons \cite{ltime,DHD}. However, there is some difference
in concrete numerical values of lifetimes obtained in different
papers. In papers \cite{DHD} we have commented on the
uncertainties in the resulting values of lifetimes related to the
values of heavy quark masses. Besides this, there is one more
uncertainty remained due to the value of light quark - diquark
wave-function at origin. Today there are two approaches to
estimate this value: 1) assuming, that this value is the same as
the value of $D$ - meson wave function at origin; 2) extracting
this value from the comparison of hyper-fine splittings in doubly
heavy and singly heavy baryons. Here we give the results of the
lifetime estimates made in the second approach, as they are the
most complete.
\begin{table}[t]
\begin{center}
\begin{tabular}{|c|c|c|c|}
\hline & $\Xi_{cc}^{++}$ & $\Xi_{cc}^{+}$ & $\Omega_{cc}^{+}$ \\
\hline $\sum c\to s$, ps$^{-1}$ & 3.104 & 3.104 & 3.104 \\ \hline
PI, ps$^{-1}$ & -0.874 & - & 0.621 \\ \hline WS, ps$^{-1}$ & - &
1.776 & - \\ \hline $\tau$, ps & 0.45 & 0.20 & 0.27 \\ \hline
\end{tabular}
\end{center}
\caption{The lifetimes of doubly charmed baryons  together with
the relative spectator and nonspectator contributions to the total
widths.} \label{cc}
\end{table}

\begin{table}[t]
\begin{center}
\begin{tabular}{|c|c|c|c|}
\hline & $\Xi_{bc}^{+}$ & $\Xi_{bc}^{0}$ & $\Omega_{bc}^{0}$ \\
\hline $\sum b\to c$, ps$^{-1}$ & 0.632 & 0.632 & 0.631 \\ \hline
$\sum c\to s$, ps$^{-}$ & 1.511 & 1.511 & 1.509 \\ \hline PI,
ps$^{-1}$ & 0.807 & 0.855 & 0.979 \\ \hline WS, ps$^{-1}$ & 0.653
& 0.795 & 1.713 \\ \hline $\tau$, ps & 0.28 & 0.26 & 0.21 \\
\hline
\end{tabular}
\end{center}
\caption{The lifetimes of $(bcq)$-baryons  together with the
relative spectator and nonspectator contributions to the total
widths.} \label{bc}
\end{table}

\begin{table}[t]
\begin{center}
\begin{tabular}{|c|c|c|c|}
\hline & $\Xi_{bb}^{0}$ & $\Xi_{bb}^{-}$ & $\Omega_{bb}^{-}$ \\
\hline $\sum b\to c$, ps$^{-1}$ & 1.254 & 1.254 & 1.254 \\ \hline
PI, ps$^{-1}$ & - & -0.0130 & -0.0100 \\ \hline WS, ps$^{-1}$ &
0.0189 & - & - \\ \hline $\tau$, ps & 0.79 & 0.80 & 0.80 \\ \hline
\end{tabular}
\end{center}
\caption{The lifetimes of $(bbq)$-baryons  together with the
relative spectator and nonspectator contributions to the total
widths.} \label{bb}
\end{table}

From Tables 1.-3. we see a sizeable contribution of nonspectator
effects to the lifetimes of doubly heavy baryons. The presence of
the latter, for example, leads to a huge difference of
$(ccq)$-baryon lifetimes.

\section{Exclusive decays in NRQCD sum rules}

In this section we review the results for exclusive decay modes of
doubly heavy baryons, obtained in the framework of three-point
NRQCD sum rules. Our consideration of form-factors, governing the
above transitions, will be restricted to the case of spin $1/2$ -
spin $1/2$ baryon transitions. We will comment on the size of spin
$1/2$ - spin $3/2$ contribution in the section with our numerical
results.

\subsection{Two point sum rules}

We start with the two-point NRQCD sum rules for corresponding
baryonic couplings. For baryons, containing two heavy quarks,
there are two distinct choices of baryonic interpolating currents:

\noindent 1) The prescription with the explicit spinor structure
of the heavy diquark from the very beginning
\begin{eqnarray}
J_{\Xi^{\prime \diamond}_{QQ^{\prime}}} &=& [Q^{iT}C\tau\gamma_5
Q^{j\prime}]q^k\varepsilon_{ijk},\nonumber\\
J_{\Xi_{QQ}^{\diamond}} &=& [Q^{iT}C\tau\mathbf{\gamma }^m
Q^j]\cdot\mathbf{\gamma}_m\gamma_5 q^k\varepsilon_{ijk},
\label{def}
\end{eqnarray}
\noindent 2) The currents, which require further symmetrization of
heavy diquark wave function
\begin{eqnarray}
J_{\Xi_{QQ}^{\diamond}} = \varepsilon^{\alpha\beta\gamma }
:(Q^{T}_{\alpha}C\gamma_5 q_{\beta })Q^{'}_{\gamma }:
\end{eqnarray}
In calculations of exclusive decay modes from three-point NRQCD
sum rules, considered below, we will use the currents of the
second type. For the benefits of this choice we refer the reader
to \cite{Onish}. The baryon couplings for both types of currents
are defined as usual
\begin{eqnarray} \langle 0|J_H|H(p)\rangle = i Z_H
u(v,M_H)e^{ip\cdot x}
\end{eqnarray}
To estimate the introduced baryonic couplings, we consider
two-point correlation function of corresponding currents
\begin{eqnarray}
\Pi(^2) &=& i\int d^4x e^{ipx}\langle 0|T\{J(x),\bar
J(0)\}|0\rangle = \nonumber \\ && \slashchar{v} F_1(p^2) +
F_2(p^2),\label{2pcor}
\end{eqnarray}
where $v$ is the four-velocity of the studied doubly heavy baryon.

\begin{figure}[ph]
\begin{center}
\begin{picture}(100,80)
\put(0,34){\epsfxsize=6cm \epsfbox{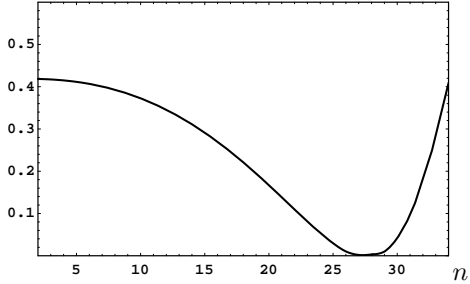}} \put(60,35){$n$}
\put(0,77){$\Delta M_{\Xi_{bc}}$, GeV}
\end{picture}
\end{center}
\vspace*{-3.5cm} \caption{The difference between the
$\Xi_{bc}$-baryon masses calculated in the NRQCD sum rules for the
formfactors $F_1$ and $F_2$ in the scheme of moments for the
spectral densities (first type of currents).} \label{mbc}
\end{figure}

\begin{figure}[th]
\begin{center}
\begin{picture}(100,80)
\put(5,30){\epsfxsize=6cm \epsfbox{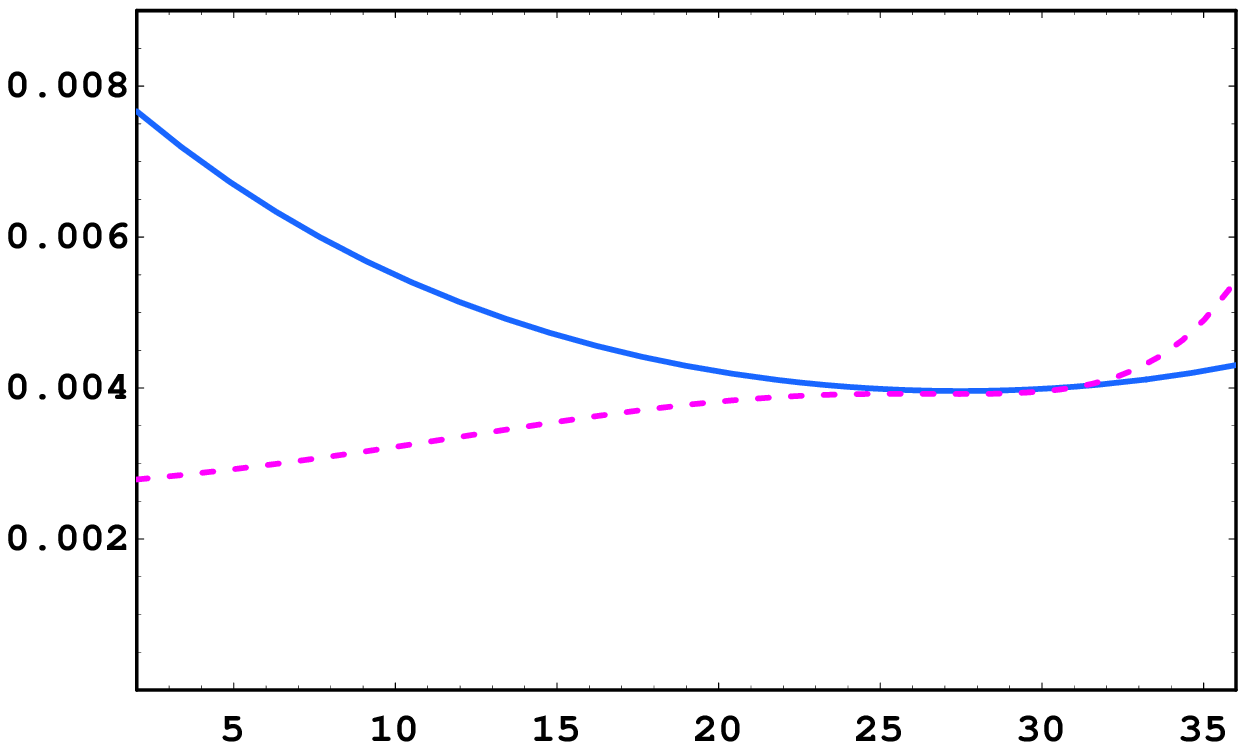}} \put(65,30){$n$}
\put(0,76){$|Z_{\Xi_{bc}}|^2$, GeV$^2$}
\end{picture}
\end{center}
\vspace*{-3.5cm} \caption{The couplings $|Z_{\Xi_{bc}}^{(1,2)}|^2$
of $\Xi_{bc}$-baryon calculated in the NRQCD sum rules for the
formfactors $F_1$ and $F_2$ in the scheme of moments for the
spectral densities (first type of currents).} \label{zbc}
\end{figure}

\begin{figure}[th]
\begin{center}
\vspace*{7cm}
\begin{picture}(100,80)
\put(5,35){\epsfxsize=6cm \epsfbox{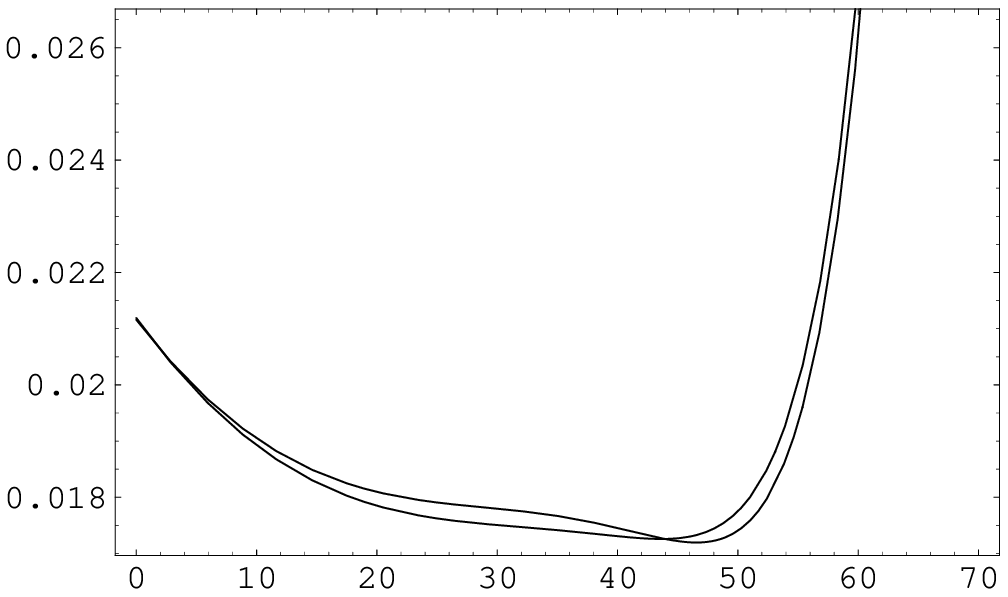}} \put(65,105){$n$}
\put(0,150){$|Z_{\Xi_{bb}}|^2$, GeV$^2$}
\end{picture}
\end{center}
\vspace*{-11cm} \caption{The couplings $|Z_{\Xi_{bb}}^{(1,2)}|^2$
of $\Xi_{bb}$-baryon calculated in the NRQCD sum rules for the
formfactors $F_1$ and $F_2$  in the scheme of moments for the
spectral densities (second type of currents).} \label{zbb}
\end{figure}

For both types of currents the NRQCD sum rules derived include
coulomb-like corrections in the system of doubly heavy diquark as
well as contributions of nonperturbative terms coming from the
quark, gluon, mixed condensates and the product of quark and gluon
condensates\cite{DHSR1,Onish}. As was shown by the authors of
\cite{QCDsr}, for the second type of currents it is difficult, in
general, to achieve a stability of sum rules predictions for the
both extracted mass and coupling of doubly heavy baryons. So, for
the second type of currents used here, we evaluate the coupling
constants only and use the masses of doubly heavy baryons,
calculated by us previously \cite{DHSR1}, as inputs.

The results of the performed analysis can be most conveniently
understood from the figures below. The plotted result for the
$\Xi_{bb}$-baryon coupling of the second type does not include the
Coulomb corrections, as the calculation of desired form-factors
for the doubly heavy baryons can be consistently performed without
accounting for Coulomb corrections either, provided we neglect
them both in the two-point and three-point sum rules\footnote{For
more details see \cite{Onish,Valera}}.

\subsection{Three-point sum rules}

\setlength{\unitlength}{1mm}
\begin{figure}[th]
\begin{center}
\vspace*{2.cm}
\begin{picture}(200,150)
\put(5,-5){\epsfxsize=6cm \epsfbox{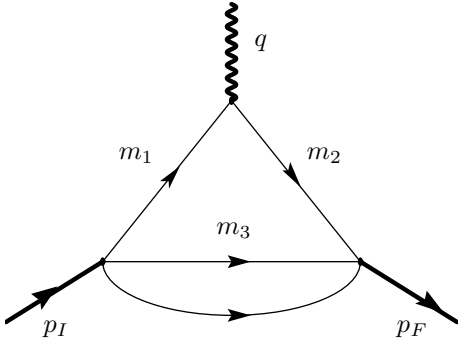}}
\put(20,150){$m_1$} \put(45,150){$m_2$} \put(33,140){$m_3$}
\put(10,127){$p_I$} \put(57,127){$p_F$} \put(38,165){$q$}
\end{picture}
\end{center}
\vspace*{-12.5cm} \caption{The diagram, corresponding to the
three-point correlation function considered in the paper.}
\label{diagram}
\end{figure}
\normalsize

Following the standard  procedure for the evaluation of
form-factors in the framework of NRQCD sum rules, we consider the
three-point correlation function
\begin{eqnarray}
\Pi_{\mu } &=& i^2\int d^4xd^4y \langle
0|T\{J_{H_F}(x)J_{\mu}(0)\bar J_{H_I} \}|0\rangle\times\nonumber
\\ && e^{i p_F\cdot x}e^{-i p_I\cdot y}
\end{eqnarray}
The theoretical expression for the three-point correlation
function can be easily calculated with the use of double
dispersion relation
\begin{eqnarray}
&& \Pi_{\mu }^{(theor)}(s_1,s_2,q^2) = \nonumber \\ &&
\frac{1}{(2\pi
)^2}\int_{m_I^2}^{\infty}ds_1\int_{m_F^2}^{\infty}ds_2\frac{\rho_{\mu
}(s_1,s_2,q^2)}{(s_1-s_1^0)(s_2-s_2^0)} + \ldots \nonumber \\
\end{eqnarray}
Saturating the channels of initial and final state hadrons  by
ground states of corresponding baryons, we have the following
phenomenological expression for the three-point correlation
function
\begin{eqnarray}
\Pi_{\mu }^{(phen)}(s_1,s_2,q^2) &=& \sum_{spins}\frac{\langle
0|J_{H_F}|H_F(p_F)\rangle}{s_2^0-M_{H_F}^2}\times\nonumber \\ &&
\langle H_F(p_F)|J_{\mu}|H_I(p_I)\rangle\times \nonumber \\ &&
\frac{\langle H_I (p_I)|\bar J_{H_I}|0\rangle }{s_1^0-M_{H_I}^2}
\end{eqnarray}
The formfactors for spin $\frac{1}{2}$ -- spin $\frac{1}{2}$
baryon transitions are modeled as following
\begin{eqnarray}
\langle H_F(p_F)|J_{\mu}|H_I(p_I)\rangle &=& \bar u(p_F) \{
\gamma_{\mu} G^V_1 + v_{\mu }^I G^V_2 + \nonumber \\ && v_{\mu }^F
G^V_3 + \gamma_5 (\gamma_{\mu} G^A_1 + \nonumber \\ && v_{\mu }^I
G^A_2 + v_{\mu }^F G^A_3 )\} u(p_I)\nonumber
\end{eqnarray}
Naively, all these six formfactors  are independent. However, the
analysis of spin symmetry relations in the limit of zero recoil
shows the semileptonic decays of doubly heavy baryons can be
described by the only universal function, an analogue of
Isgur-Wise function\cite{Onish,Lozano}.

So, now we in position, where to obtain estimates on semileptonic
or nonleptonic transitions under hypothesis of factorization
\cite{fact} we should calculate the only universal function.

The calculation of spectral densities is straightforward with the
use of Cutkosky rules for quark propagators \cite{Cutk}.  The
results for the trace of correlation function with $v^I_{\mu }$
are

\noindent 1) heavy to heavy underlying quark transition
\begin{eqnarray}
\rho^{pert} &=&
\int_{m_3^2}^{(\sqrt{s_1}-m_1)^2}\frac{6}{(2\pi)^4}\frac{m_1 m_2
(k^2-m_3^2)^2}{k^2(\lambda (s_1,s_2,q^2))^{1/2}} dk^2\nonumber \\
&& \\ \rho^{\bar qq} &=& -\frac{4}{(2\pi )^2}\frac{m_1 m_2 m_3
\sqrt{s_1}}{(m_1+m_3)(\lambda (s_1,s_2,q^2))^{1/2}}\langle\bar
qq\rangle \nonumber \\ && \\ \cos\theta &=& \frac{m_2}{|\vec
p_2||\vec k|} ( \sqrt{s1}-p_{20}+(m_2-m_1)\times\nonumber \\ &&
(1-\frac{|\vec k|^2}{2m_1m_2})+\frac{|\vec p_2|^2}{2 m_2})
\end{eqnarray}

\noindent 2) heavy to light underlying quark transition
\begin{eqnarray}
\rho^{pert} &=& \int_{m_3^2}^{(\sqrt{s_1}-m_1)^2}\frac{3 F_1}{4
(2\pi)^4}\frac{(k^2-m_3^2)^2}{k^2(\lambda (s_1,s_2,q^2))^{1/2}}
dk^2 \nonumber \\ && \\ \rho^{\bar qq} &=& -\frac{m_1 m_3
\sqrt{s_1} F_2}{2 (m_1+m_3)(\lambda
(s_1,s_2,q^2))^{1/2}}\langle\bar qq\rangle \nonumber \\ &&
\end{eqnarray}
\begin{eqnarray}
\cos\theta &=& \frac{1}{2|\vec p_2||\vec k|} (
2p_{20}(\sqrt{s_1}-m_1-\frac{|\vec k|^2}{2m_1})-s_2 \nonumber \\
&& -(\sqrt{s1-m_1})^2+\frac{\sqrt{s_1}|\vec k|^2}{ m_1}+m_2^2)
\end{eqnarray}
where
\begin{eqnarray}
F_1 &=&
\frac{2}{\sqrt{s_1}}(m_1^2-q^2+2m_2\sqrt{s_1}+s_2-k^2)\nonumber
\\ && \\ F_2 &=& F_1|_{k^2\to m_3^2}
\end{eqnarray}
The notations in the above expressions should be clear from Fig.
6. Having derived theoretical expressions for the three-point
correlation function, we may proceed now with the evaluation of
form-factors. In numerical estimates we will use the Borel scheme
for the form-factor extraction and so, below we give the formula
determining the universal Isgur-Wise function for the semileptonic
decays of doubly heavy baryons
\begin{eqnarray}
\xi^{IW}(q^2) &=& \frac{1}{(2\pi )^2}\frac{1}{8 M_I M_F Z_I Z_F
}\nonumber \\ &&\int_{(m_1+m_3)^2}^{s_I^{th}}
\int_{(m_1+m_2)^2}^{s_F^{th}} \rho(s_I,s_F,q^2)ds_Ids_F \nonumber
\\ && \times\exp (-\frac{s_I-M_I^2}{B_I^2})\exp
(-\frac{s_F-M_F^2}{B_F^2}),\nonumber \\
\end{eqnarray}
where $B_I$ and $B_F$ are the Borel parameters in the initial and
final state channels.

\subsection{Numerical estimates}

The analysis of NRQCD sum rules in the Borel scheme gives us the
estimates of the value of Isgur-Wise (IW) function at zero recoil
for different types of spin $1/2$ - spin $1/2$ transitions betweem
doubly heavy baryons, shown in Table 4. For the sake of
comparison, we also provide here the estimates of the values of
IW-function at zero recoil performed by us in the framework of
potential models, which can be also found in Table 4. We see that
within the errors of the sum rule method (15\%) the obtained
results are very close to each other.

\begin{table}[th]
\begin{center}
\begin{tabular}{|c|c|c|}
\hline\hline Mode & $\xi (1)$ SR &  $\xi (1)$ PM \\ \hline
$\Xi_{bb}\to \Xi_{bc}$ & 0.85 & 0.91 \\\hline $\Xi_{bc}\to
\Xi_{cc}$ & 0.91 & 0.99 \\\hline $\Xi_{bc}\to \Xi_{bs}$ & 0.9 &
0.99
\\\hline $\Xi_{cc}\to \Xi_{cs}$ & 0.99 & 1. \\\hline\hline
\end{tabular}
\end{center}
\caption{The normalization of Isgur-Wise function for different
baryon transitions at zero recoil.}
\end{table}

In Fig.7 we have plotted the dependence of the normalization of
IW-function on the Borel parameters of initial and final state
baryons in the case of $\Xi_{bb}\to \Xi_{bc}$ transition.

\setlength{\unitlength}{1mm}
\begin{figure}[th]
\begin{center}
\vspace*{1.5cm}
\begin{picture}(200,150)
\put(5,15){\epsfxsize=6cm \epsfbox{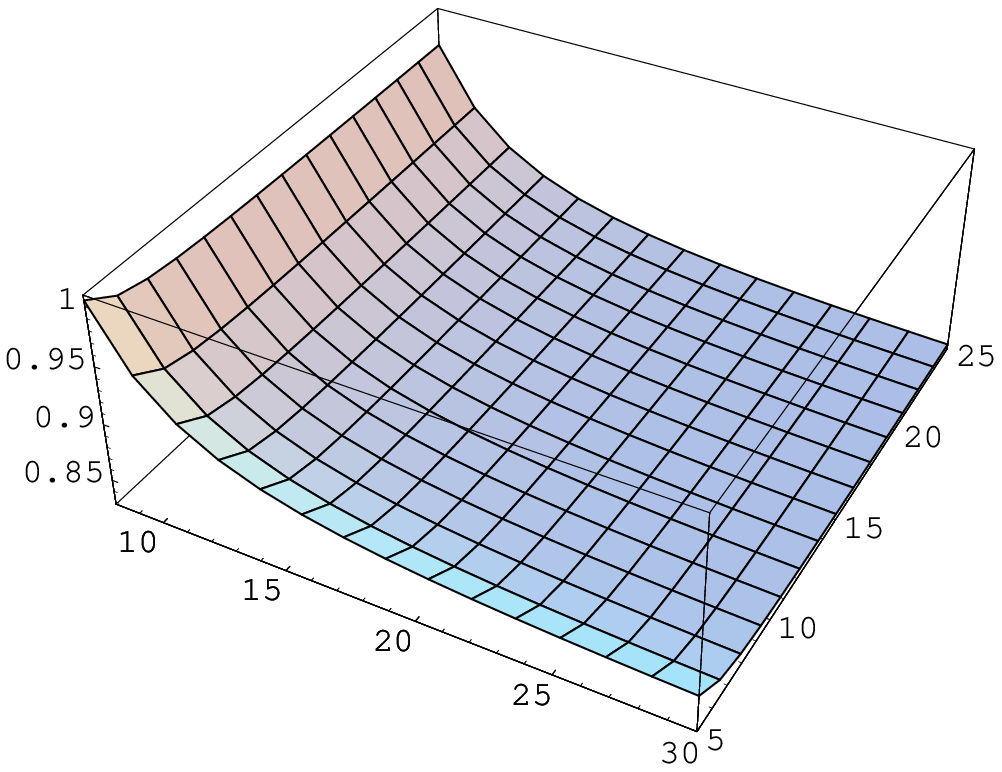}} \put(5,148){$\xi
(1)$} \put(25,115){$B_I$} \put(60,125){$B_F$}
\end{picture}
\end{center}
\vspace*{-11.5cm} \caption{The value of $\xi (1)$ for the
transition $\Xi_{bb}^{\diamond}\to \Xi_{bc}^{\diamond}$ as
function of Borel parameters in the $s_I$ and $s_F$ channels.}
\label{3bb}
\end{figure}
\normalsize

Next, to obtain the dependence of formfactors on the square of
momentum transfer we exploit the pole resonance model. So, for the
IW-function we have the following expression:
\begin{equation}
\xi^{IW} (q^2) = \xi_0\frac{1}{1-\frac{q^2}{m_{pole}^2}},
\end{equation}
with
\begin{eqnarray}
m_{pole} &=& 6.3 \mbox{~~~GeV for the~~}  b\to c
\mbox{~~transitions}\nonumber \\ m_{pole} &=& 1.85 \mbox{~~GeV for
the ~~} c\to s \mbox{~~transitions}.\nonumber
\end{eqnarray}

With the obtained estimates for the form-factors, we can easily
obtain the predictions for the semileptonic and some nonleptonic
decay modes of doubly heavy baryons. The results of such estimates
can be found in Table 2.

\begin{table}[th]
\begin{center}
\begin{tabular}{|c|c|c|c|}
\hline\hline Mode & Br (\%) &  Mode & Br (\%) \\ \hline
$\Xi_{bb}^{\diamond}\to \Xi_{bc}^{\diamond}l\bar\nu_l$ & 14.9 &
$\Xi_{bc}^{+}\to \Xi_{cc}^{++}l\bar\nu_l$ & 4.9 \\\hline
$\Xi_{bc}^{0}\to \Xi_{cc}^{+}l\bar\nu_l$ & 4.6 & $\Xi_{bc}^{+}\to
\Xi_{bs}^{0}\bar l\nu_l$ & 4.4 \\\hline $\Xi_{bc}^{0}\to
\Xi_{bs}^{-}\bar l\nu_l$ & 4.1 & $\Xi_{cc}^{++}\to
\Xi_{cs}^{+}\bar l\nu_l$ & 16.8 \\\hline $\Xi_{cc}^{+}\to
\Xi_{cs}^{0}\bar l\nu_l$ & 7.5 & $\Xi_{bb}^{\diamond}\to
\Xi_{bc}^{\diamond}\pi^{-}$ & 2.2
\\\hline $\Xi_{bb}^{\diamond}\to \Xi_{bc}^{\diamond}\rho^{-}$ &
5.7 & $\Xi_{bc}^{+}\to \Xi_{cc}^{++}\pi^{-}$ & 0.7 \\\hline
$\Xi_{bc}^{0}\to \Xi_{cc}^{+}\pi^{-}$ & 0.7 & $\Xi_{bc}^{+}\to
\Xi_{cc}^{++}\rho^{-}$ & 1.9 \\\hline $\Xi_{bc}^{0}\to
\Xi_{cc}^{+}\rho^{-}$ & 1.7 & $\Xi_{bc}^{+}\to
\Xi_{bs}^{0}\pi^{+}$ & 7.7 \\\hline $\Xi_{bc}^{0}\to
\Xi_{bs}^{-}\pi^{+}$ & 7.1 & $\Xi_{bc}^{+}\to
\Xi_{bs}^{0}\rho^{+}$ & 21.7 \\\hline $\Xi_{bc}^{0}\to
\Xi_{bs}^{-}\rho^{+}$ & 20.1 & $\Xi_{cc}^{++}\to
\Xi_{cs}^{+}\pi^{+}$ & 15.7 \\\hline $\Xi_{cc}^{+}\to
\Xi_{cs}^{0}\pi^{+}$ & 11.2 & $\Xi_{cc}^{++}\to
\Xi_{cs}^{+}\rho^{+}$ & 46.8 \\\hline $\Xi_{cc}^{+}\to
\Xi_{cs}^{0}\rho^{+}$ & 33.6 &  & \\\hline\hline
\end{tabular}
\end{center}
\caption{Branching ratios for the different decay modes of doubly
heavy baryons. }
\end{table}

To calculate the branching ratios for exclusive decay modes we
used the values of doubly heavy baryon lifetimes, calculated by us
previously \cite{DHD}.  The values, presented in Table 2 already
include the contribution of spin $1/2$-spin $3/2$ decay channels.
To estimate the latter we have used the results of \cite{Lozano},
where the contribution of these channels was calculated for the
case of $\Xi_{bc}\to \Xi_{cc}+l\bar\nu$ baryon transition, and
assumed, that, according to superflavor symmetry, it constitutes
30 \% from the contribution of corresponding spin $1/2$-spin $1/2$
transitions for all transitions between doubly heavy baryons. In
calculations of $\Xi_{bb}^{\diamond }$ and $\Xi_{cc}^{\diamond }$
- baryon decay modes we have taken into account a factor 2 due to
Pauli principle for the identical heavy quarks in the initial
channel. In the case of $\Xi_{bc}^{\diamond }\to
\Xi_{cc}^{\diamond '} X$-baryon transition the same factor comes
from the positive Pauli interference of the $c$-quark, being a
product of $b$-quark decay, with the $c$-quark from the initial
baryon. Here, we also would like to mention, that for the
$\Xi_{bc}$-baryon decays the mentioned positive Pauli interference
contribution is dominant compared to other nonspectator
contributions\footnote{Here we use the results of OPE analysis for
the inclusive decay modes of doubly heavy baryons
\cite{ltime,DHD}}, so we do not introduce other corrections here.
However, in the case of $\Xi_{cc}^{++}\to \Xi_{cs}^{+} X$- baryon
transition the negative Pauli interference plays the dominant role
and thus should be accounted for explicitly. From the previously
done OPE analysis for doubly heavy baryon lifetimes
\cite{ltime,DHD} we conclude that the corresponding correction
factor in this case is $0.62$. We would like also give a small
comment on our notations. The $\Xi_{Qs}^{\diamond}$ in Table 2
stays for the sum of $\Xi_{Q}^{\diamond}$ and $\Xi_{Q}^{\diamond
'}$ decay channels. The obtained results are in agreement with the
previous estimates of $\Xi_{bc}$-baryon exclusive decay modes
\cite{Lozano} and with the results of OPE analysis
\cite{ltime,DHD} for inclusive decay modes.

\section{Conclusion}

In this paper we have made a short review on the inclusive and
exclusive decay modes of doubly heavy baryons. The results on the
lifetimes of doubly heavy baryons as well as estimates of
semileptonic, pion and $\rho $-meson decay modes are given.

This work was in part supported by the Russian Foundation for
Basic Research, grants 99-02-16558 and 00-15-96645, by
International Center of Fundamental Physics in Moscow,
International Science Foundation and INTAS-RFBR-95I1300.


\begin{thebibliography}{**}
\bibitem{cdf-bc}
{ F. Abe et al.}, CDF Collaboration, Phys. Rev. Lett. {\bf 81},
2432 (1998), Phys. Rev. {\bf D58}, 112004 (1998).
\bibitem{bc-rev}
{ S.S.Gershtein, V.V.Kiselev, A.K.Likhoded, A.V.Tkabladze,
A.V.Berezh\-noy, A.I.Onish\-chen\-ko}, Talk given at 4th
International Workshop on Progress in Heavy Quark Physics,
Rostock, Germany, 20-22 Sept. 1997, IHEP 98-22 [hep-ph/9803433];\\
{ S.S.Gershtein, V.V.Kiselev, A.K.Likhoded, A.V.Tkabladze}, Phys.
Usp. {\bf 38}, 1 (1995) [Usp. Fiz. Nauk {\bf 165}, 3 (1995)];\\ {
S.S.Gershtein et al.}, Phys. Rev. {\bf D51}, 3613 (1995).
\bibitem{prod}
{ A.V.Berezhnoy, V.V.Kiselev, A.K.Likhoded, A.I.Onishchenko},
Phys. Rev. D57 (1998) 4385;\\ { A.V.Berezhnoy, V.V.Kiselev,
A.K.Likhoded}, Z.Phys. {\bf A356}, 89 (1996), Phys. Atom. Nucl.
{\bf 59}, 870 (1996) [Yad. Fiz. {\bf 59}, 909 (1996)];\\ {
S.P.Baranov}, Phys. Rev. D 56, 3046 (1997);\\ { V.V.Kiselev, A.K.
Likhoded, M.V. Shevlyagin}, Phys. Lett. B332, 411 (1994);\\ {
A.Falk et al.}, Phys. Rev. D49, 555 (1994);\\ V.V.Kiselev,
A.E.Kovalsky, preprint hep-ph/9908321.
\bibitem{ltime}
{ V.V.Kiselev, A.K.Likhoded, A.I.Onishchenko}, Phys. Rev. {\bf
D60}, 014007 (1999), Phys.Atom.Nucl. {\bf 62} (1999) 1940,
Yad.Fiz. {\bf 62} (1999) 2095; \\ { V.V.Kiselev, A.K.Likhoded,
A.I.Onishchenko}, preprint DESY 98-212 (1999) [hep-ph/ 9901224],
to appear in Eur.Phys.J. {\bf C} ;\\ { B.Guberina, B.Melic,
H.Stefancic}, Eur. Phys. J. {\bf C9}, 213 (1999).
\bibitem{DHD}
{A.I.Onishchenko}, preprint hep-ph/9912424; \\ {A.K.Likhoded,
A.I.Onishchenko}, preprint hep-ph/9912425.
\bibitem{pot}
{ S.S.Gershtein, V.V.Kiselev, A.K.Likhoded, A.I.Onishchenko},
preprint IHEP 98-66 (1998) [hep-ph/9811212], Heavy Ion Phys {\bf
9}, 133 (1999); [hep-ph/9807375] Mod. Phys. Lett. {\bf A14}, 135
(1999);\\ { D.Ebert, R.N.Faustov, V.O.Galkin, A.P.Martynenko,
V.A.Saleev}, Z. Phys. C76 (1997) 111;\\ { J.G.K$\ddot o$rner,
M.Kr$\ddot a$mer, D.Pirjol}, Prog. Part. Nucl. Phys. 33 (1994)
787;\\ { R.Roncaglia, D.B.Lichtenberg, E.Predazzi}, Phys. Rev. D52
(1995) 1722.
\bibitem{SVZ}
{ M.A.Shifman, A.I.Vainshtein, V.I.Zakharov}, Nucl. Phys. {\bf
B147}, 385 (1979);\\ { L.J.Reinders, H.R.Rubinstein, S.Yazaki},
Phys. Rep. {\bf 127}, 1 (1985).
\bibitem{QCDsr}
{ E.Bagan, M.Chabab, S.Narison}, Phys. Lett. B306 (1993) 350;\\ {
E.Bagan at al.}, Z. Phys. {\bf C64}, 57 (1994).
\bibitem{DHSR1}
{V.V.Kiselev, A.I.Onishchenko}, preprint hep-ph/9909337, to appear
in Nucl.Phys. {\bf B}.
\bibitem{DHSR2}
{V.V.Kiselev, A.E.Kovalsky}, preprint hep-ph/0005019.
\bibitem{Lozano}
{M.A.Sanchis-Lozano}, Nucl.Phys. {\bf B440} (1995) 251.
\bibitem{Guo}
{X.-H.Guo, H.-Y.Jin, X.-Q.Li}, Phys.Rev {\bf D58} (1998), 114007.
\bibitem{Onish}{A.I.Onishchenko}, preprint hep-ph/0006271.
\bibitem{Smilga}
{ A.Smilga}, Yad. Phys. {\bf 35}, 473 (1982).
\bibitem{Cutk}
{ R.E.Cutkosky}, J. Math. Phys. {\bf 1}, 429 (1960).
\bibitem{Valera}
{ V.V.Kiselev, A.V.Tkabladze}, Phys. Rev. {\bf D48},(1993), 5208;
\\ { V.V.Kiselev, A.K.Likhoded, A.I.Onishchenko}, Nucl.Phys.
{\bf B569}, (2000), 473; { V.V.Kiselev, A.E.Kovalsky,
A.K.Likhoded}, hep-ph/0002127
\bibitem{fact}
M.Dugan and B.Grinstein, Phys.Lett. {\bf B255} (1991) 583;\\
M.A.Shifman, Nucl.Phys. {\bf B388} (1992) 346; \\ B.Blok,
M.Shifman, Nucl.Phys. {\bf 389} (1993) 534. \bibitem{SS} H.Georgi
and M.B.Wise, Phys.Lett. {\bf B243} (1990) 279;\\ C.D.Carone,
Phys.Lett. {\bf B253} (1991) 408;\\ M.J.Savage and M.B.Wise,
Phys.Lett. {\bf B248} (1990) 177.
\end{thebibliography}
\end{document}